\documentclass[twocolumn,prd,nofootinbib]{revtex4}
\usepackage{graphicx}
\usepackage{epstopdf}
\usepackage{epsfig}

\newcommand{\npix}{N_{\rm pix}}
\newcommand{\nbdy}{N_{\rm bdy}}
\newcommand{\nsrc}{N_{\rm src}}
\newcommand{\beq}{\begin{equation}}
\newcommand{\eeq}{\end{equation}}

\usepackage{graphicx}
\usepackage{amsmath}

\begin{document}
\title{Efficient decomposition of cosmic microwave background
polarization maps into pure E, pure B, and ambiguous components}
\author{Emory F. Bunn}
\email{ebunn@richmond.edu}
\affiliation{Physics Department, University of Richmond, Richmond, VA  23173}

\begin{abstract}
Separation of the B component of a cosmic microwave background (CMB)
polarization map from the much larger E component is an essential step
in CMB polarimetry.  For a map with incomplete sky coverage, 
this separation is necessarily hampered by the presence of ``ambiguous''
modes which could be either E or B modes.  I present an efficient pixel-space
algorithm for removing the ambiguous modes and separating the map
into ``pure'' E and B components.  The method, which
works for arbitrary geometries, does not involve generating
a complete basis of such modes and scales the cube of the number of pixels
on the boundary of the map.
\end{abstract}

\maketitle

\section{Introduction}

A great deal of attention in cosmology is focused on attempts to characterize the polarization
of the cosmic microwave background (CMB) radiation.  Multiple experiments have detected CMB polarization \cite{dasi,cbi,wmap,boomerang,capmap,quad}.  The Planck Surveyor \cite{planck} will provide all-sky polarization maps at many frequencies, and many other experiments are in development.  
Many in the cosmology community regard the quest for high-quality CMB polarization maps as an extremely high priority \cite{weissreport,whitepaper}.  CMB polarization maps can potentially provide a great deal
of information about our Universe, complementing the wealth of information provided by
CMB temperature anisotropy. 

Although polarization maps have multiple uses, much of the interest in CMB polarization stems from the prediction that a stochastic background of gravitational waves, prouced during an inflationary epoch, may be visible in polarization maps. Detection of this background would be of revolutionary importance, providing direct confirmation of inflation as well as a measurement of the inflation energy scale.

The possibility of detecting this gravitational wave background is a consequence of the fact that a CMB polarization map can be regarded as a sum of two components, a scalar $E$ component and a pseudoscalar $B$ component \cite{kkslett,selzal,zalsel,kks}.  To linear order in perturbation theory, ``ordinary'' scalar perturbations populate only the $E$ component.  The tensorial gravitational waves, on the other hand, populate both $E$ and $B$ components.  Because the tensor component is known to be smaller in amplitude than the scalar component, detection of it requires a channel that is free of scalar contributions.  $B$-component polarization may provide this channel.

 The $B$ component is predicted to be at least an order of magnitude smaller than the $E$ component on all angular scales.  It is therefore important to make sure that any detection of $B$-type polarization is free from contamination from the $E$ component.  Such contamination can be caused by systematic errors, of course \cite{hhz,bunnsys,shimon,odea}, but even in an error-free map one must worry about mixing of $E$ and $B$ components in the data analysis process \cite{bunndetect}.  For a map with complete sky coverage, the two components can be perfectly separated in the spherical harmonic domain, but for a map with only partial sky coverage care must be taken to separate components in a leakage-free way \cite{lewis,LCT,BZT,bunnmnproc,bunncmbpolproc}.
 
One way to think about this concern is in the language of the ``pure''
and ``ambiguous'' components of a polarization map \cite{BZT}.  A
``pure $E$'' (resp.\ $B$) mode on the incomplete sky is one that can
only have been produced by $E$-type (resp.\ $B$-type) polarization,
while an ambiguous mode is one that could have been produced by either
component. To be specific, a (not necessarily pure) 
$E$ (resp.\ $B$) mode ${\bf P}$ 
is one that satisfies a 
certain differential relationship ${\bf D}_B^\dag\cdot{\bf P}=0$
(resp.\ ${\bf D}_E^\dag\cdot{\bf P}=0$). Explicit forms for the
differential operators ${\bf D}_{E,B}$ are given in Section \ref{sec:sep},
and further details may be found in
refs. \cite{LCT,BZT}.
An ambiguous mode is one
that satisfies both conditions simultaneously. There are no
modes satisfying both conditions on the complete sphere, but there
on any domain consisting of only part of the 
sphere.\footnote{The analogy between spin-two polarization fields and
spin-one vector fields is helpful here. On a two-dimensional surface
without boundary, such as a sphere, any smooth vector field can be
uniquely decomposed into curl-free and divergence free components.
On a surface with boundary, however, the decomposition is not
unique due to the existence of ``ambiguous'' 
vector fields that are both curl-free and divergence-free, such
as $x\hat{\bf x}-y\hat{\bf y}$, defined over a subset of the plane.
}

``Pure'' $E$ (resp.\ $B$) modes are defined to be orthogonal to \textit{all}
$B$ (resp.\ $E$) modes, including the ambiguous modes. Explicit examples
of pure and ambiguous modes may be found in ref.\ \cite{BZT}.

On the incomplete sky, the decomposition into $E$ and $B$ components
is not unique, since there is no way to decide where to put the ambiguous
modes. One way to illustrate this is to imagine working with a square
patch of sky in the flat-sky approximation. The
$E$-$B$ decomposition is trivial in Fourier space, so one could perform the 
decomposition by Fourier transforming, decomposing, and transforming
back. Now imagine performing this set of operations after padding the 
observed map out to a larger size, inserting any values you like in
the unobserved region. Different $E$/$B$ decompositions will result
depending on how the padding is performed, all of which will match
the data over the observed region. If all one has access to is the 
observed patch, there is no way to tell which if any of these is
the ``real'' $E$/$B$ decomposition.

Although the $E/B$ decomposition is not unique for a partial sky map,
the decomposition into \textit{pure} $E$, \textit{pure} $B$, and ambiguous
components is unique \cite{BZT}.
In practice, if
the $E$ component dominates over the $B$ component as expected, the
ambiguous modes will mostly contain information about $E$ modes, and a
robust detection of $B$-type polarization must therefore be sought in the pure
$B$ component.  One consequence of this loss of $B$ information to
ambiguous modes is a shift in the optimum tradeoff between sensitivity
and sky coverage \cite{bunndetect}: the optimum sky coverage for a
$B$-type experiment is larger than would be found by a straightforward
``Knox formula'' \cite{knoxformula}.
 
The information loss due to ambiguous modes has two sources: incomplete sky coverage and pixelization \cite{BZT}.  Pixelization ambiguity is essentially a consequence of aliasing of Fourier modes (working in the flat-sky approximation for simplicity).  In a map with pixel size $L_{\rm pix}$, Fourier modes with wavevector components greater than the Nyquist frequency $k_{\rm Ny}=2\pi/L_{\rm pix}$ are aliased to modes of lower frequency.  In the process, the wave vector $\vec k$ is mapped to a new vector with, generically, a completely different direction.  The decomposition of a Fourier mode into $E$ and $B$ components depends entirely on the direction of $\vec k$, so aliasing thoroughly scrambles $E$ and $B$ modes.  There is only one way to avoid this: one must pixelize finely, pushing the Nyquist frequency to a level where beam suppression makes aliasing negligible.  

If a data set has been pixelized too coarsely, resulting in significant aliasing, no data analysis method can undo the $E$-$B$ mixing.  Therefore, in this paper, I will assume that the data have been sufficiently finely pixelized (relative to the beam size) to control pixelization-induced $E$-$B$ mixing to an acceptable level.  The method described in this paper is aimed at removing the incomplete-sky-induced ambiguous component.

Of course, no matter how fine the pixelization, the noise will not be smooth on the pixel scale.  Separate tests are therefore required to make sure that the decomposition described herein is well-behaved with respect to noise in the data.  Section \ref{sec:tests} discusses such tests.
 
The original method for decomposition into pure E, pure B, and ambiguous components (hereinafter referred to as an E/B/A decomposition) involved the construction of an orthornormal basis of such modes by solution of an eigenvector problem of dimension $2N_{\rm pix}$, the number of pixels in the data set.  Such an operation is computationally extremely expensive for large data sets.  This paper will present an alternative algorithm that is far more efficient.  

Even on an incomplete sky, it is easy to separate a polarization map into $E$ and $B$ components, if one does not worry about the purity of those components: one simply performs the operation in Fourier space (if the flat-sky approximation is appropriate) or spherical harmonic space; in either case the separation can be done mode by mode.  The difficulty comes in projecting out the ambiguous component from each of these components.  The ambiguous component of each map is determined by data on the boundary of the map, so it is natural to seek methods of finding it and projecting it out by examining only pixels near the boundary of the observed region.  This paper will present one such method.  As we will see, such methods can be far more efficient than the na\"ive method of finding a complete set of normal modes.

In principle, in order to perform power spectrum estimation (the primary goal of a CMB polarization experiment), it is not necessary to perform any E/B/A (or indeed E/B) separation at all.  For any given choice of $E$ and $B$ power spectra, one can in principle compute the likelihood function $L(C_l^E,C_l^B)$ and use it  to draw confidence intervals or Bayesian credible regions in power spectrum space.  If this analysis excludes $C_l^B=0$, then $B$-type power has been detected.  For large data sets, where the full likelihood is too expensive to compute, other methods such as the pseudo-$C_l$ method have been generalized from temperature anisotropy to polarization data \cite{smithpseudocl,smith2,smithzal,grain,challchon}.  Such methods achieve near-optimal power spectrum estimates without the need to perform an explicit E/B/A decomposition.  There may well be other data analysis methods that do not involve worrying about an E/B/A decomposition.  For example, one might analyze interferometric data entirely in visibility space \cite{parkng,hamilton,charlassier}, without ever constructing a real-space map at all.

Nonetheless, E/B/A separation of polarization data sets will be useful
for several reasons.  Although the power spectra are the primary
quantities to be measured in a CMB data set, they are not the end of
the story; real-space maps are necessary for a variety of
applications.  Probably the most important will be tests for
foreground contamination, which may be easier to do via real-space
cross-correlation with foreground templates. Use of the
lensing $B$-mode signal to constrain cosmological parameters (e.g.,
\cite{kapknoxsong,smithhukap,lesgourgues}) depends on real-space
$B$-component information, rather than just the power spectra.
Searches for
non-Gaussianity and departures from statistical isotropy also go
beyond the power spectrum.  A real-space picture of the pure $B$
component will be an important ``sanity check'' to make sure that the
detected $B$ component looks qualitatively as expected.  Last but
certainly not least, people (both scientists and the broader public)
will find it much easier to believe that $B$ modes have really been
detected if there is an actual map they can be shown.

Other methods have been proposed for performing $E$-$B$ separation on
incomplete sky maps \cite{kimnaselsky,zhao,caofang,bowyer}.  These methods
may prove extremely useful, but the method I propose herein differs
from them in significant ways: it allows reconstruction of the actual
polarization map components (i.e., the observables $Q,U$) as opposed
to a scalar derivative of these quantities, and it does so by solving
the relevant differential equation for the ambiguous modes, rather
than by making a heuristic approximation that the ambiguous modes can
be regarded as confined to the boundary of the map.

The remainder of this paper is organized as follows.  Section \ref{sec:sep} reviews the formalism behind pure and ambiguous components and lays out the schematic recipe for the E/B/A separation.  Section \ref{sec:implement} describes the technical details of the implementation.  Section \ref{sec:tests} describes some tests of the algorithm, and Section \ref{sec:discuss} contains a brief discussion.
  
\section{Separation into pure and ambiguous modes}
\label{sec:sep}

\subsection{Review of formalism}

We begin with a review of some useful relations involving $E$ and $B$ modes.  The reader wishing further detail can see, e.g., refs.~\cite{LCT,BZT}. For simplicity, we work in the flat-sky approximation.  Section \ref{sec:discuss} discusses the generalization to the spherical sky.

Let $Q(\vec r)$ and $U(\vec r)$ be the Stokes parameters as functions
of position $\vec r$ on the sky.  We will group them together 
into a vector field
\beq
\mathbf{p}=\begin{pmatrix}Q\\U\end{pmatrix}.
\eeq
Here and throughout, arrows denote spatial vectors, while boldface
denotes vectors in more general vector spaces.  In particular, $\mathbf{p}$
is not a vector in position space -- i.e., it transforms with
spin two rather than one.  

For the present we neglect pixelization effects and allow ourselves to take derivatives.  A polarization field is an E mode if it
satisfies a second-order differential relation
\beq
\mathbf{D}_B^\dag\cdot\mathbf{p}=0,
\eeq
and is a B mode if it satisfies
\beq
\mathbf{D}_E^\dag\cdot\mathbf{p}=0.
\eeq
In the flat-sky approximation, the differential operators in these equations
can be written
\begin{eqnarray}
\mathbf{D}_E&=&\begin{pmatrix}\partial_x^2-\partial_y^2\\
2\partial_x\partial_y\end{pmatrix}\\
\mathbf{D}_B&=&\begin{pmatrix}-2\partial_x\partial_y\\
\partial_x^2-\partial_y^2\end{pmatrix}
\end{eqnarray}
These two operators are the spin-2 analogues of the divergence
and curl respectively.
They satisfy the following useful relations:
\begin{eqnarray}
\mathbf{D}_E^\dag\cdot\mathbf{D}_B=\mathbf{D}_B^\dag\cdot\mathbf{D}_E&=&0,
\label{eq:dedb}\\
\mathbf{D}_E^\dag\cdot\mathbf{D}_E=\mathbf{D}_B^\dag\cdot\mathbf{D}_B&=&\nabla^4
\equiv(\nabla^2)^2.
\label{eq:dededbdb}
\end{eqnarray}
When working on the sphere rather than the plane, the operators $\mathbf{D}_{E,B}$ take on a 
more complicated form, and the bilaplacian
$\nabla^4$ is replaced by $\nabla^2(\nabla^2+2)$.

Just as with curl- and divergence-free vector fields, we can express $E$ and $B$ modes in terms of
potentials: any polarization field $\mathbf{p}_E$ 
that satisfies the $E$ mode condition
can be written as the derivative of a potential $\psi_E$, and
similarly for any $B$ mode $\mathbf{p}_B$:
\begin{eqnarray}
\mathbf{p}_E&=&\mathbf{D}_E\psi_E,\label{eq:efrompsi}\\
\mathbf{p}_B&=&\mathbf{D}_B\psi_B.
\end{eqnarray}

An ambiguous mode, by definition, is one that simultaneously
satisfies the requirements of both $E$ and $B$ modes.  We can construct
such modes by choosing a biharmonic potential $\psi$, i.e., one with
\beq
\nabla^4\psi=0.
\eeq
Then both $\mathbf{D}_E\psi$ and $\mathbf{D}_B\psi$ will be ambiguous
modes: equations (\ref{eq:dedb}) and (\ref{eq:dededbdb}), along
with the biharmonicity condition, imply that both $\mathbf{D}_B^\dag$ and
$\mathbf{D}_E^\dag$ yield zero when applied to these fields.

A ``pure'' $E$ mode is defined to be one that is orthogonal, over the
observed region, to all $B$ modes (including the ambiguous modes).
Since it is an $E$ mode, a pure $E$
mode can always be derived from a potential $\psi_E$ via
equation (\ref{eq:efrompsi}), and in order to be pure the potential must
satisfy both Dirchlet and Neumann boundary conditions on the
boundary $\partial\Omega$ of the observed region:
\beq
\left.\psi_E\right|_{\partial\Omega}=
\left.\vec n\cdot\nabla\psi_E\right|_{\partial\Omega}=0,
\eeq
where $\vec n$ is normal to the boundary.
There is a unique biharmonic function satisfying a given
set of Dirichlet and Neumann boundary conditions.  Subtracting off
the biharmonic function that matches the boundary conditions
of $\psi_E$ gives
a unique way to ``purify'' a given E mode.

\subsection{Schematic recipe for E/B/A decomposition}

Suppose that we have a polarization field $\mathbf{p}$ observed
over a region $\Omega$.  If we ignore pixelization issues and assume that
we can differentiate, then we can decompose the field into pure E, pure B, and 
ambiguous components as follows:
\begin{enumerate}
\item Decompose $\mathbf{p}$ into E and B components without
worrying about purity, specifically by finding a pair of potentials
$\psi_E,\psi_B$ such that 
\beq
\mathbf{p}=\mathbf{D}_E\psi_E+\mathbf{D}_B\psi_B.
\eeq
\item Find functions $\alpha_E,\alpha_B$ that are biharmonic, i.e.,
\beq
\nabla^4\alpha_{X}=0
\eeq
(where $X$ is either $E$ or $B$) 
and that match the potentials on the boundaries:
\begin{eqnarray}
\left.\alpha_{X}\right|_{\partial\Omega}&=&
\left.\psi_{X}\right|_{\partial\Omega},\\
\left.\vec n\cdot\nabla\alpha_{X}\right|_{\partial\Omega}&=&
\left.\vec n\cdot\nabla\psi_{X}\right|_{\partial\Omega}.
\end{eqnarray}
\item ``Purify'' the potentials by defining
\beq
\psi_{pE}=\psi_E-\alpha_E,\qquad
\psi_{pB}=\psi_B-\alpha_B.
\eeq
\item Apply the differential operators to obtain the pure E, pure B,
and ambiguous polarization fields:
\begin{eqnarray}
\mathbf{p}_{pE}&=&\mathbf{D}_E\psi_{pE}\label{eq:pureefinal}\\
\mathbf{p}_{pB}&=&\mathbf{D}_B\psi_{pB}\\
\mathbf{p}_a&=&\mathbf{D}_E\alpha_E+\mathbf{D}_B\alpha_B.\label{eq:ambfinal}
\end{eqnarray}
\end{enumerate}

For pixelized data, we can follow a similar procedure, with 
appropriately defined differential operators.  Of course, we then have to test whether the discretization has introduced significant errors.

\section{Implementation}
\label{sec:implement}

The procedure described in the previous section is straightforward
in principle, but to implement it numerically on a discretized
grid requires some care.  In this section I will describe
this procedure in more detail.  I begin with a summary of the key
points.  The reader with sufficient patience can then proceed to examine
the technical details in the rest of the section.

Before embarking on the above recipe, we need to embed our data
in a rectangular grid, regarded as satisfying periodic boundary conditions, suitable for
discrete Fourier transforms.  We will then 
apply the various differential operators in the Fourier
domain.\footnote{One can
instead work with discretizations of the differential
operators involving nearest neighbors \cite{bowyer}.  Such methods are inexact
even when applied to low-frequency Fourier modes and 
lead to non-negligible E/B mixing even for modes significantly
below the Nyquist frequency.}   
Since the observed
data will cover only a portion of the grid,\footnote{Even if the data
lie on a rectangular grid, it is necessary to pad it out to a larger
grid to avoid artifacts due to periodic boundary conditions.}
we must extend it into the unobserved region.  This extension
must be smooth in order to avoid artifacts near the boundaries
when we differentiate.

The Fourier-space differential operators are exact for functions that are band limited below the Nyquist frequency.  We assume that the pixelization is fine enough, compared to the experimental beam, that the intrinsic signal has negligible power above the Nyquist frequency.  The smooth extension must be designed so that the extended data are also smooth enough on the pixel scale to have power spectra that become negligible well before the Nyquist scale. I describe one method for smoothly extending
the data in Section \ref{sec:extend}.

Once the data have been extended, we can begin the
decomposition.  Step 1 is straightforward to implement
in the Fourier domain, since the E/B decomposition can be 
done mode by mode  in Fourier space
(Section \ref{sec:decomp}).

Next we must find biharmonic functions satisfying the 
required boundary conditions.  This can be done, e.g., by
relaxation methods, but another method
appears to be more efficient.  We seek a function $\alpha$
such that $\nabla^4\alpha=0$ within the observed region $\Omega$
and $\alpha$ satisfies certain boundary conditions on $\partial\Omega$.
We can identify a set of ``source'' points $S$ outside of the observed
region on which $\nabla^4\alpha$ is allowed to be nonzero.
By solving a linear system, we find the values of $\nabla^4\alpha$ on $S$
such that, when the inverse bilaplacian operator $\nabla^{-4}$
is applied, the resulting  $\alpha$ satisfies the required boundary conditions.
The inverse bilaplacian is a simple
convolution with a fixed kernel and is efficient to apply in the Fourier
domain.
Section \ref{sec:biharm} supplies details.

This procedure completes step 3 in our recipe, and step 4 is then
trivially implemented in the Fourier domain, as described in the very brief Section \ref{sec:laststep}.

\subsection{Smooth extension of data}
\label{sec:extend}
Suppose that we have a data set 
$\mathbf{p}_{\rm obs}$ defined on a subset of 
the pixels in our grid.  We need to extend it smoothly
to a 
field $\mathbf{p}$ that covers the entire grid and 
matches $\mathbf{p}_{\rm obs}$ where
data exist.

There are no doubt many ways to achieve this goal.  One choice would be to create a realization of a Gaussian random process, constrained to match the observed data.  Smoothness of the extension can then be enforced by assigning a steeply declining power spectrum to the random process.  

There is a well-established process \cite{hoffmanribak1,hoffmanribak2} for generating constrained realizations of Gaussian random fields.  The process involves two steps: first, the mean field of the constrained random process is found, and then random fluctuations are generated about that mean field.  For our present purposes, we can simply stop after the first step: the mean field matches the constraints and is in fact smoother than the typical realization, so it is better for our purpose.


\begin{figure*}[t]
\begin{minipage}[b]{2.25in}
\epsfig{width=2.25in,figure=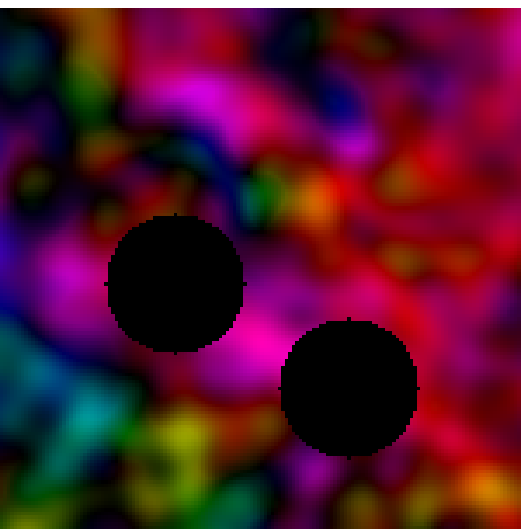}

\epsfig{width=0.75in,figure=colorwheel.eps}
\end{minipage}\hspace{0.5in}
\begin{minipage}[b]{3in}
\epsfig{width=3in,figure=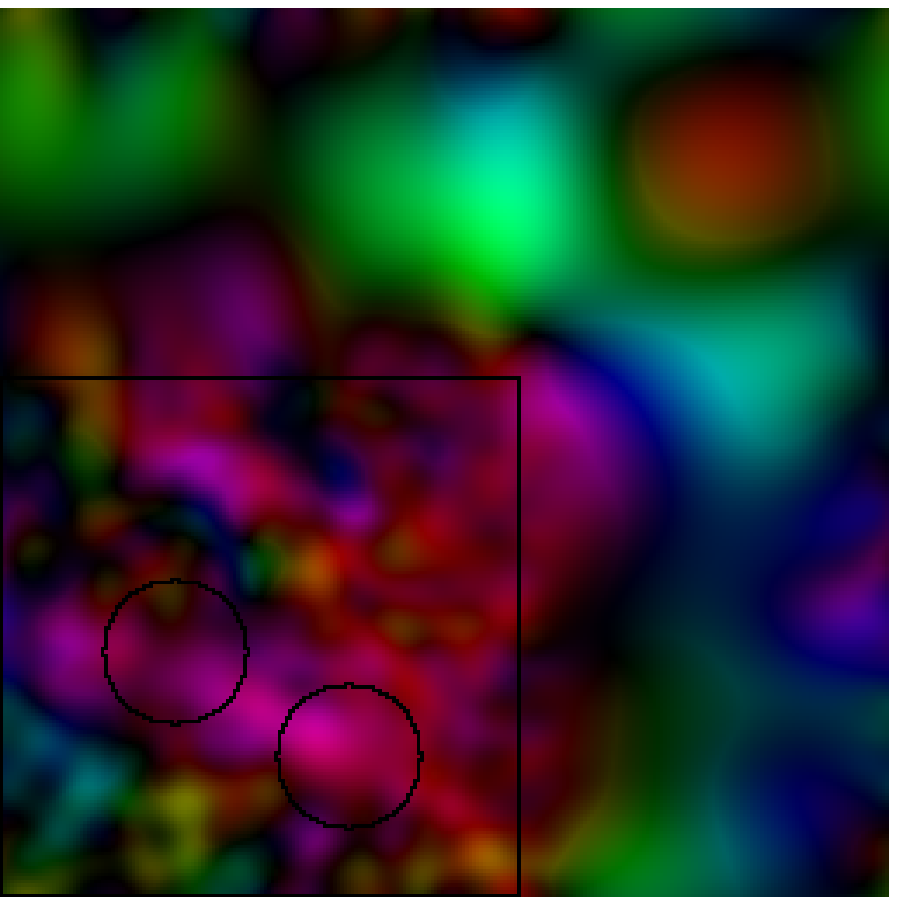}
\end{minipage}

\caption{Illustration of maximum-likelihood extension.  The
upper left panel shows a simulated data set, observed over a square region
with two holes in it.  The direction of polarization is indicated
by hue and the polarization amplitude by value
in the HSV color system.  The disc in the lower left illustrates this scheme,
with polarization amplitude increasing with distance from the center
and direction varying with angle.  
The data
were simulated with power spectrum $P_E(k)\propto k^{-2}$ with
no B modes, and smoothed with a Gaussian beam of width $\sigma=4$ pixels. 
The right panel shows the mean-field Gaussian extension of the data.
The
extension was calculated using a power spectrum $P(k)\propto k^{-4}$, with
$P_B=0.01P_E$, and constrained to match the data over a 3-pixel boundary
region.}
\label{fig:extend}
\end{figure*}

We now describe this process in detail.
Consider a Gaussian random process that generates a polarization
map $\mathbf{g}=(Q,U)$, characterized by $E$ and $B$ mode power spectra
$P_E(k),P_B(k)$.  Suppose that the values of $\mathbf{g}$ are constrained
to match $\mathbf{p}_{\rm obs}$ over the observed region, or at
least over a subset of the observed region lying near the boundary.
In practice, it is sufficient to choose to impose the constraint
only on a band of pixels consisting of the $T_{\rm ext}$ nearest neighbors of
boundary pixels, for some reasonably small $T_{\rm ext}$.
We wish
to determine the most probable realization of $\mathbf{g}$ on the 
rest of the pixels.

It is important to emphasize that the resulting $\mathbf{g}$ 
is not supposed to be the ``real'' polarization map extrapolated
into the unobserved region.  It is merely an artificial field
designed solely to join smoothly onto the observed field.  
The power spectra $P_E,P_B$ do not have to match those of the real data;
in fact, it is better if they are very steeply declining, so that the
Gaussian random process will strongly favor smooth functions.

Let the number of pixels in the entire grid be $N_{\rm pix}$ and the number 
of constraint pixels be $N_{\rm cons}$.  Imagine writing the 
values of the Gaussian random field over the entire grid
in a $2N_{\rm pix}$-dimensional vector $\mathbf{G}$.  Some
subset of these points are the $2N_{\rm cons}$ constraint
values.  These are listed in a $2N_{\rm cons}$-dimensional
vector $\mathbf{C}$.

Let 
\beq
\Xi_{ij}=\langle G_iG_j\rangle
\eeq
be the correlation between any two elements of the (unconstrained)
Gaussian random
field. We use these values to define two matrices: $\mathbf{\Xi}^{cc}$
is the $2N_{\rm cons}\times 2N_{\rm cons}$ dimensional matrix 
giving the correlations between constraint points, and $\mathbf{\Xi}^{gc}$
is the $2N_{\rm pix}\times 2N_{\rm cons}$ dimensional matrix
giving the correlations between arbitrary points and constraint
points.
Then 
the mean field, or most probable, value of $\mathbf{G}$ is
\beq
\mathbf{G}= \mathbf{\Xi}^{gc}\cdot (\mathbf{\Xi}^{cc})^{-1}\cdot\mathbf{C}.
\label{eq:meanfield}
\eeq

Although this procedure sounds cumbersome, it is quite simple to 
apply.  The covariance matrix elements $\Xi_{ij}$ are related in
simple ways to the Fourier transforms of the power spectra $P_E,P_B$.
Moreover, they depend only on the vector separation between
two pixels, which means that the multiplication by the matrix
$\mathbf{\Xi}^{gc}$ in equation (\ref{eq:meanfield}) is a convolution
that can be done in time $N_{\rm pix}\ln N_{\rm pix}$.
Calculation of the vector $(\mathbf{\Xi}^{cc})^{-1}\cdot\mathbf{C}$,
on the other hand, requires time  $O(N_{\rm cons}^3)$.  For a reasonably convex map with a ``nice'' boundary, the number of boundary pixels is of order the square root of the total number of pixels, in which case this is equivalent to $O(N_{\rm obs}^{3/2})$.

\begin{figure*}
\epsfig{width=1in,figure=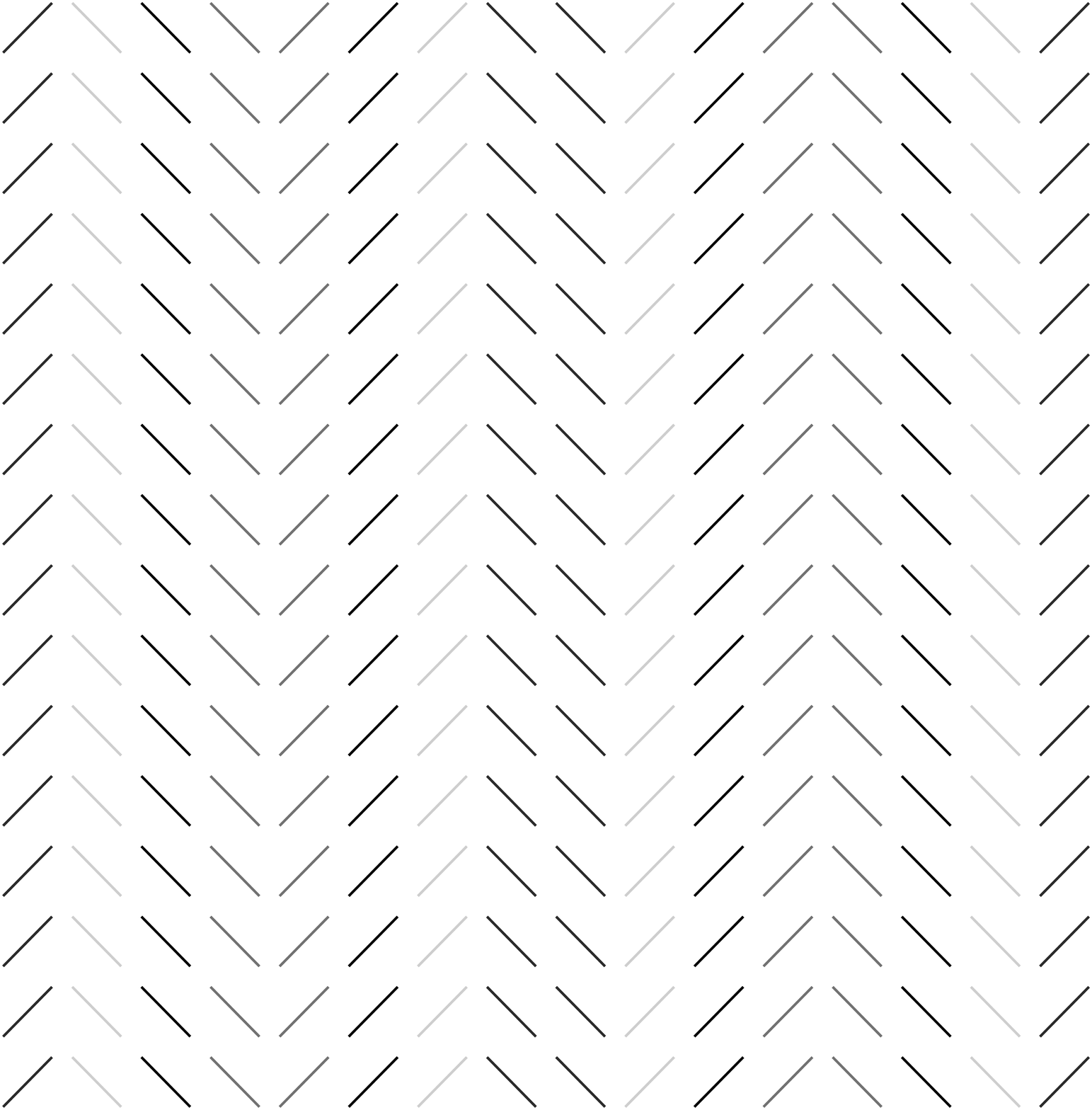}\hskip 0.5in
\epsfig{width=2in,figure=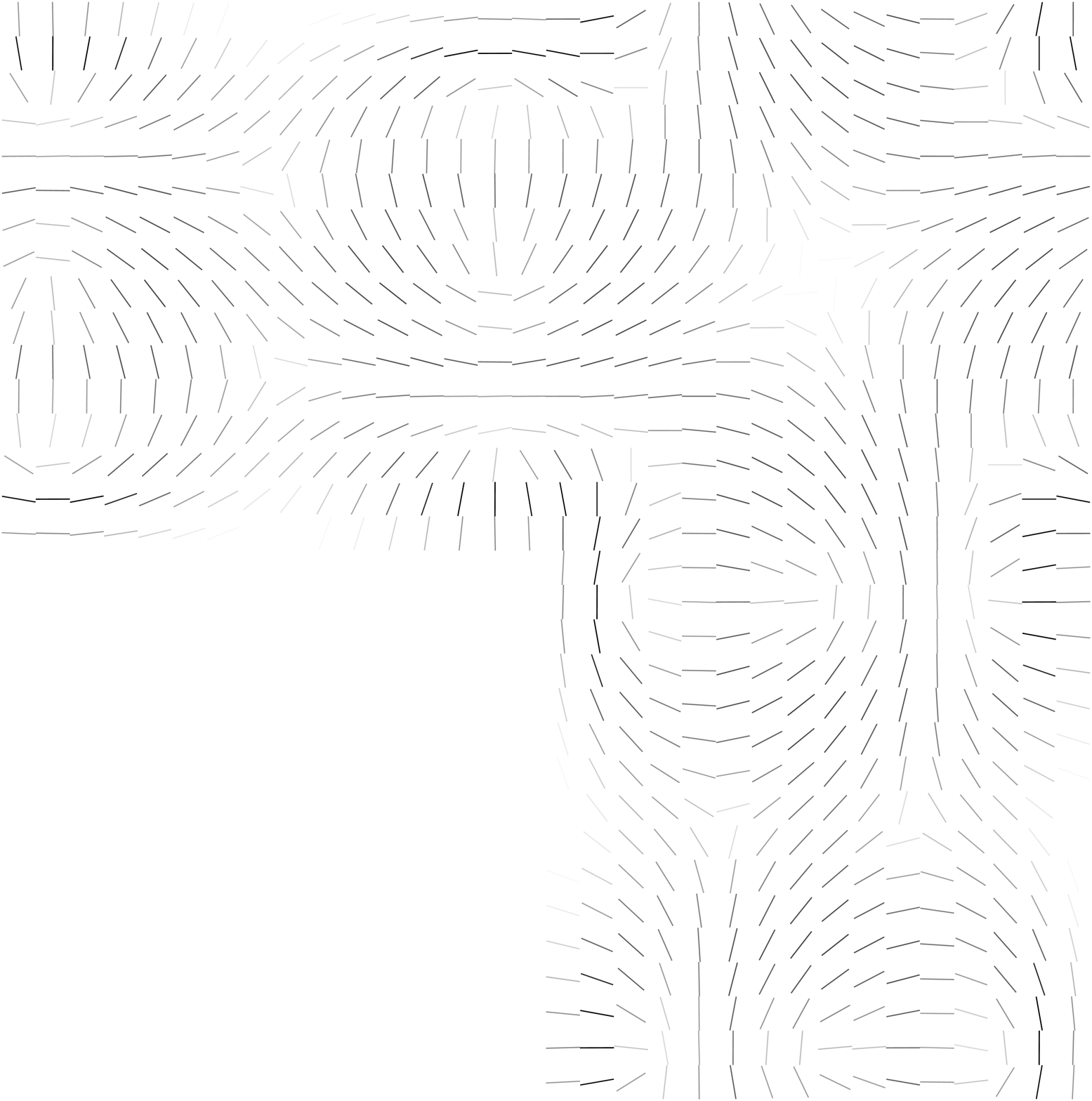}
\epsfig{width=2in,figure=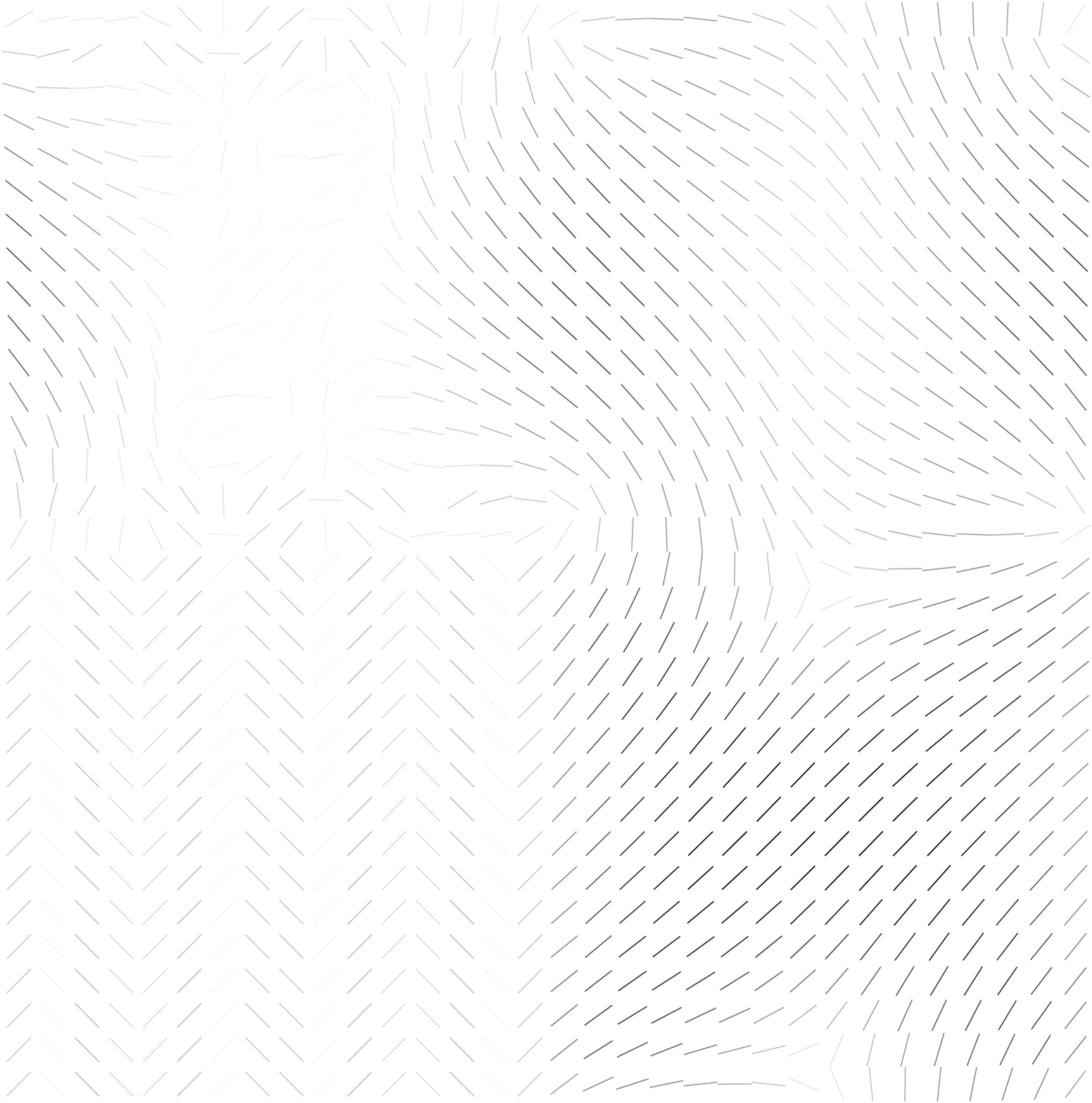}

\caption{Numerical instability in extending maps.  The original map
(left panel) is a single B-type Fourier mode.  It
was extended using the mean-field method, with results
shown in the center and right panels.  The center
panel shows the result of the extension when the Gaussian
random process had no B power.  The result is an attempt to match the input $B$ mode using only $E$ modes, leading to numerical instability: the maximum polarization level
in the extended region is 590 times that of the original map, so
that the original map (lower left quadrant) is invisible.  The 
right panel shows the result obtained when $P_B$ is taken to be $0.01P_E$.
In this case, the peak polarization level is 4.3 times that in the extended
region.}
\label{fig:extendinstability}
\end{figure*}

Note that the computation of the $\mathbf{\Xi}$ matrices, as well as 
the Cholesky decomposition required on $\mathbf{\Xi}^{cc}$ depend only
on the geometry of the system and are thus precomputable.

Figure \ref{fig:extend} shows an example of this process. The ``observed''
region is a $150\times 150$ pixel patch with two circular holes 10 pixels
in radius.  The data consists of an E-type polarization
field with a power spectrum $P_E(k)\propto k^{-2}$.  (The simulation
was made on a much larger $1200\times 1200$ grid and truncated, so that 
the observed grid does not satisfy periodic boundary conditions.)
The observed region has been extended to a $256\times 256$ grid,
including filling in the holes, following the procedure above.
The extension was constrained to match the observed data over a boundary
layer 3 pixels thick.  The Gaussian random process assumed for the extension
had both $P_E$ and $P_B$ proportional to $k^{-4}$ and $P_B=.01P_E$.

The extension
smoothly joins onto the observed region and satisfies periodic boundary
conditions, so that any further computations involving discrete
Fourier transforms will be free of artifacts from discontinuities.
Note that the extension becomes featureless far from the observed region.
This simply reflects the fact that, when the constraints 
are unimportant, the most
probable value of a Gaussian random process is uniformity.

Some choices must be made in applying this method, 
namely the thickness of the boundary layer on which to force
matching, and the power spectra
for both E and B modes in the Gaussian random process assumed
for the extension.
The choice of boundary
layer thickness is governed by the desire for computational efficiency:
if it does not lead to excessive computation time, there is no reason
not to constrain on the entire observed region.  However, since
the purpose of the constraint is simply to insure smoothness of the
extension, this is not necessary: numerical tests indicate that 
a boundary layer of a few pixels is sufficient.

The choice of power spectrum does not seem to make very much difference
either, as long as it is a strongly decreasing function of $k$.  I adopt a power-law power spectrum $P(k)\propto k^{-n_s}$, typically with $n_s=4$.   The method
can be applied with different power spectra for E and B modes.  I recommend
setting $P_E$ to be much larger than $P_B$, so that the extension will
be nearly all E modes.  This has the result that artifacts resulting
from the extension will infect the pure E map more than the pure B map
and thus do less damage.  However, $P_B$ should not be 
taken to be zero: if the data set contains actual B modes, 
trying to satisfying the constraints with only E modes can lead
to numerical instability as shown in Figure 2.

Section \ref{sec:tests} discusses 
some numerical tests on these parameters, showing that the final
pure and ambiguous maps are insensitive to the choices made over
a broad range.

\subsection{Initial decomposition}
\label{sec:decomp}

After generating a smooth extension of the data onto a rectangular
grid with periodic boundary conditions, decomposing the 
data into (impure) E and B components is trivial.  We take discrete
Fourier transforms of both $Q$ and $U$ and perform the decomposition
mode by mode in the Fourier plane.  For a given vector $\vec k$,
making an angle $\alpha$ with the $x$ axis, an E mode would satisfy
\begin{equation}
\begin{pmatrix}\tilde Q_E\\ \tilde U_E\end{pmatrix}=\tilde E
\begin{pmatrix}\cos 2\alpha\\ \sin 2\alpha\end{pmatrix}
\end{equation} 
for
some ($\vec k$-dependent) scalar $\tilde E$, and similarly,
a B mode
would satify 
\begin{equation}
\begin{pmatrix}\tilde Q_B\\ \tilde U_B\end{pmatrix}=\tilde B
\begin{pmatrix}-\sin 2\alpha\\ \cos 2\alpha\end{pmatrix}.
\end{equation} 
We add these expressions, set the sum equal to $(\tilde Q,\tilde U)$,
and solve:
\begin{eqnarray}
\tilde E&=&\tilde Q\cos 2\alpha+\tilde U\sin 2\alpha,\label{eq:etilde}\\
\tilde B&=&-\tilde Q\sin 2\alpha+\tilde U\cos 2\alpha.\label{eq:btilde}
\end{eqnarray}
The fields $E$ and $B$ are the laplacians of the corresponding
potentials $\psi_E,\psi_B$, so in Fourier space,
\beq
\tilde \psi_E=-k^{-2}\tilde E,\quad
\tilde \psi_B=-k^{-2}\tilde B.
\eeq

The monopole ($\vec k=0$) mode cannot be treated in this way, of course; it must be removed from the map and treated as an ambiguous mode.  In addition, 
modes for which either $k_x$ or $k_y$ equals the Nyquist frequency
must be treated with care: for such modes, we do not know the sign
of one component of $k$, and hence do not know the quadrant of $\alpha$.
This means that the terms proportional to $\sin 2\alpha$
in equations (\ref{eq:etilde}) and (\ref{eq:btilde}) have unknown
sign.  In the numerical implementation of this step, I separate
these pieces and place them along with the monopole in the ambiguous component at the
end of the process.  Of course, the philosophy underlying the
entire approach in this paper is that the Nyquist length is well below
the smoothing scale, in which case this is always a minor consideration.

\subsection{Solving the bilaplacian equation}
\label{sec:biharm}

We now consider the ``purification'' of $\psi_E$ and $\psi_B$.  As we have
seen, this step requires finding a biharmonic function $\alpha$
for each potential $\psi$
whose value and first derivative match the potential on the boundary.  
For efficiency's sake, we wish to avoid method that require solution
of linear systems of order $\npix$ dimensions.

One efficient method is to find
$\beta=\nabla^4\alpha$ and then apply the inverse bilaplacian
operator.  We know that $\beta=0$ in the observed region, but 
it can (and indeed must) be nonzero for some unobserved 
pixels.  We choose a set of $\nsrc$ ``source pixels,'' all lying
in the unobserved region, and allow $\beta$ to be nonzero
only on these pixels.  The boundary conditions are then
a collection of $2\nbdy$ linear equations to be solved for $\nsrc$ unknowns.
Generically, we expect solutions to exist if $\nsrc\ge 2\nbdy$.

Naturally, the solution we find will depend on the choice of source
points, and even for a given set of source points, there will
typically be multiple solutions.  We might hope that all such
solutions would coincide within the observed region, because of the
uniqueness theorem for biharmonic functions.  Unfortunately, this is
not the case: the uniqueness theorem applies to functions whose
boundary values and derivatives are specified at all boundary points,
but we are imposing only a finite, discrete set of boundary conditions
at the pixel locations.  We must hope, therefore, that a judicious
choice of source points and of solution to the linear system will
yield a good approximation to the ``correct'' solution that we seek.
In addition, of course, we must adopt criteria to judge whether we have
succeeded.

\begin{figure}
\epsfig{width=3in,figure=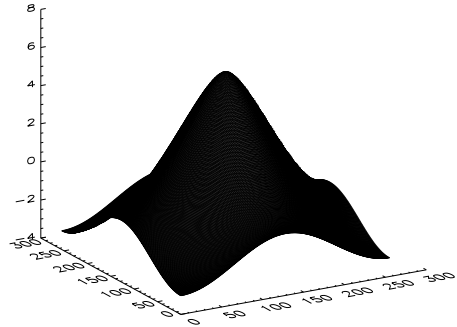}
\caption{Convolution kernel for the inverse bilaplacian.  This graph
shows the function $\alpha=\nabla^{-4}\beta$ in the case where 
$\beta$ is nonzero at only one pixel, lying at the center
of the figure.  For arbitrary $\beta$, $\alpha$ is obtained
by convolution with this kernel.}
\label{fig:invbikernel}
\end{figure}

It is plausible that the best choice of source points would be those
lying near the boundary of the observed region.  One way to see this
is to note that the relation between $\beta$ and $\alpha$ is simply
convolution with a fixed kernel: $\alpha=\nabla^{-4}\beta$ in real
space, so that $\tilde\alpha=k^{-4}\tilde\beta$ in Fourier
space.\footnote{We take the operator $\nabla^{-4}$ to have no
  monopole: $\tilde\alpha(0)=0$.}  The real-space convolution kernel
is shown in Figure \ref{fig:invbikernel}.
The kernel peaks at the source point and decays gradually.  If we try
to satisfy the boundary conditions using source points that are far
away from the boundary, the sources will have to be quite large.
Moreover, since each faraway source point will populate all of the
boundary points to comparable levels, delicate cancellations of source
points will be required.  However,  source points lying near the boundary will
primarily populate their neighboring boundary points, 
which might plausibly lead to a more stable numerical system.

Based on this heuristic reasoning, I chose source points to lie
in a boundary layer around the observed region, with the thickness $T_{\rm src}$
of the layer a free parameter.  As the tests in the next section will
show, this choice worked well.   The primary errors arose in pixels
right at the boundary.  I therefore generalized the approach to
consider source pixels that were offset from the boundary, i.e., those
whose distance $d$ (in pixels) from the boundary lay in the
interval $\Delta_{\rm src}\le d<\Delta_{\rm src}+T_{\rm src}$, where the offset $\Delta_{\rm src}$ and
the thickness $T_{\rm src}$ are both adjustable parameters.

The equations to be solved are in general underdetermined.
One way to choose a solution in this case is to solve the system
via singular value decomposition, which leads to the minimum-norm
solution to a linear system.  As we will see in the
next section, a better solution is often
obtained by retaining only some of the singular values in 
solving the system.  The number of values to retain, $N_{\rm sing}$, is thus
an additional adjustable parameter.

For a given choice of source points, the solution is
straightforward to implement.  The convolution kernel $K$ for the
inverse bilaplacian is found in the pixel domain in $O(\npix\ln\npix)$
time.  If we then represent the linear system to be solved
as a $2\nbdy\times\nsrc$ matrix $M$, then the matrix elements
for the first $\nbdy$ rows (corresponding to the Dirichlet boundary conditions)
are of the form $M_{ij}=K(\vec p_i-\vec q_j)$, where $\vec p_i$
is the location of the $i$th boundary pixel and $\vec q_j$ is
the location of the $j$th source pixel.  We can similarly 
compute the vector-valued kernel $\vec K_g$ for the gradient
of the inverse bilaplacian,  
$\vec \nabla(\nabla^{-4})$.
If $\hat n_i$ represents a unit normal to the boundary at the $i$th
boundary pixel,\footnote{One way to define this vector is to
compute the derivative of the mask, which consists of 1 for observed
pixels and 0 for unobserved pixels.  This derivative, evaluated
at boundary pixels, points in the normal direction.  Once normalized,
it can be used for $\hat n$.  In fact, as long as $\hat n$ 
is not tangent to the boundary, the results do not depend strongly
on its direction.  The reason is that the tangential derivative
of $\alpha$ is already constrained due to the Dirichlet boundary condition,
so the derivative in any linearly independent direction serves to
constrain the normal derivative.
} then the lower half of the matrix $M$  (i.e., the rows corresponding to the Neumann boundary conditions)
consists of elements of the form $\hat n_i \cdot\vec K_g(\vec p_i-\vec q_j)$.

\begingroup\squeezetable
\begin{table}
\begin{center}
\begin{tabular}{|c|c|c|}
\hline
\textbf{Extension} &Power spectrum index & $n_s=4$\\
\textbf{to unobserved} & B to E power ratio& $P_B/P_E=0.01$\\
\textbf{region}&Boundary thickness (pixels)& $T_{\rm ext}=3$\\
\noalign{\hrule}
\textbf{Solving for}& Source layer thickness (pixels) & $T_{\rm src}=10$\\
\textbf{ biharmonic} & Source layer offset (pixels)& $\Delta_{\rm src}=1$\\
\textbf{functions}& Singular values retained & $N_{\rm sing}=600$\\
\noalign{\hrule}
\end{tabular}
\caption{Fiducial values for adjustable parameters in the E/B/A decomposition.  The first three are used in extending
the data into the unobserved region (Section \ref{sec:extend}), and the last three are in solving 
the boundary-value problem for the ambiguous modes (Section \ref{sec:biharm}).}
\label{table:params}
\end{center}
\end{table}
\endgroup

\begin{figure*}
\epsfig{width=6in,figure=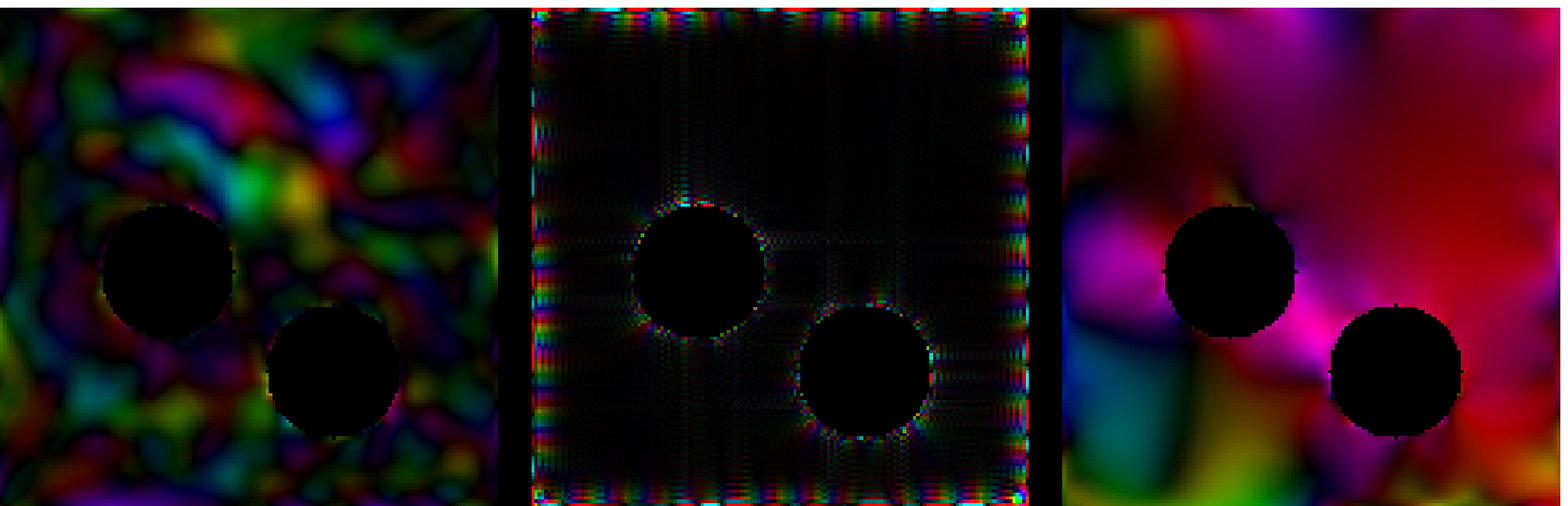}
\vskip 0.1in
\epsfig{width=6in,figure=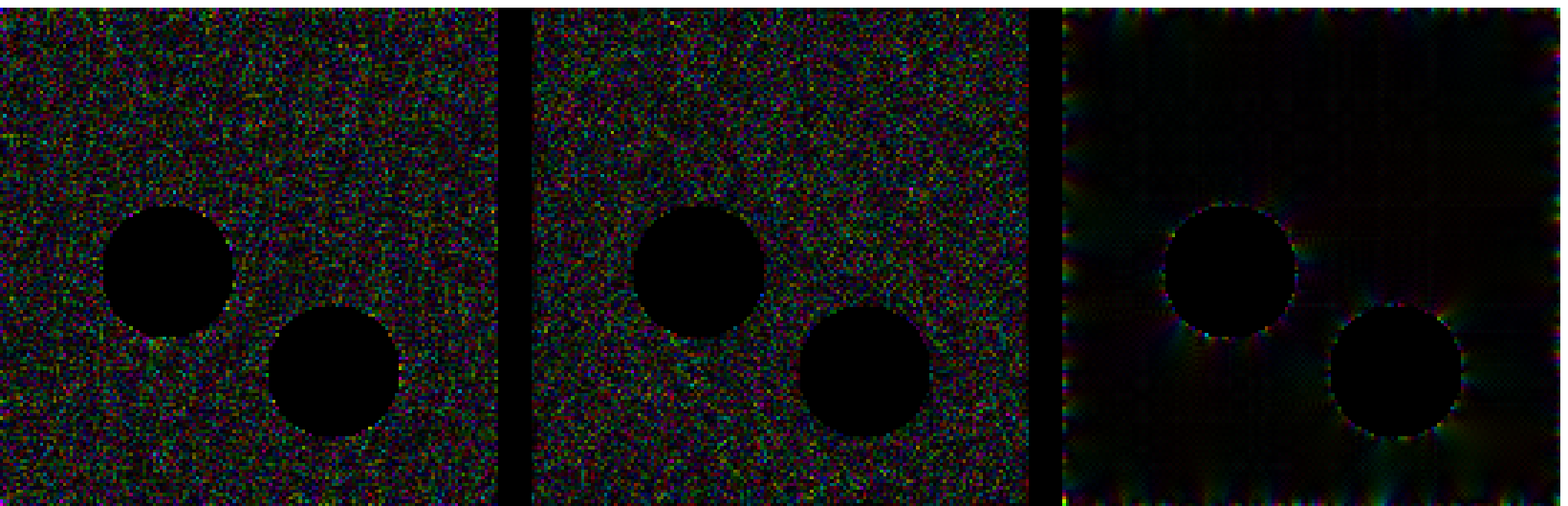}

\caption{E/B/A decomposition.  (a) The top panel shows the decomposition of the simulated map
in Figure \ref{fig:extend}.  The pure E, pure B, and ambiguous components are shown from
left to right.  The amplitude of the pure B component has been increased by a factor $10^4$.
(b) The bottom panel shows the E/B/A decomposition of a pure white noise map.  In this case,
the B component is not increased in amplitude.  In both cases, the fiducial parameters
in Table \ref{table:params} were used.}
\label{fig:decomptest}
\end{figure*}

The most time-consuming step is the singular value decomposition,
requiring $O(\nbdy^2\nsrc)\sim O(T_{\rm src}\nbdy^3)$ time.  Note, though,
that this step depends only on the pixel geometry and can be precomputed.

Once the decomposition is performed,
solving the linear system for any particular data set requires
time $O(\nsrc\nbdy)$.  Using the result to populate $\alpha$ over
the entire
oberved region is done with a convolution, requiring $O(\npix\ln\npix)$
time.

\subsection{Constructing the pure and ambiguous maps.}
\label{sec:laststep}
Once the pure and ambiguous potentials have been found, 
it is straightforward to construct the final polarization maps
by applying equations (\ref{eq:pureefinal}-\ref{eq:ambfinal})
in the Fourier domain.  
There is only one minor technical
note. In the first step, we removed both the monopole and part of 
the Nyquist-frequency modes before passing from the original map to
the potentials.  We should add these terms
into the ambiguous component at the end.

\section{Tests}
\label{sec:tests}

If we had continuously-sampled data and hence could take derivatives,
the E/B/A decomposition would be unique and exact.  The numerical
method described in the previous section reduces to the exact decomposition
in the limit where the pixelization becomes infinitely fine, but for pixelized
data it is only approximate.  

The Fourier-based
derivative operators are exact for band-limited functions but approximate for functions with power above the Nyquist frequency.  For (noise-free) data that are smoothed with a beam that is significantly larger than the pixel size, the intrinsic signal can be regarded as band-limited, to a good approximation.  So can the Gaussian process that generates the extension.  As long as the constraints imposed are sufficient to make these two maps join together smoothly, we expect, to a good approximation, to be able to regard the entire extended map as band-limited and hence differentiable in Fourier space.

Even in this case, the decomposition procedure is approximate, because the boundary conditions used to find the ambiguous modes are imposed only on a discrete set of boundary points, not continuously.  Heuristically, we expect this to pose a problem chiefly on small scales, close to the pixel scale.  As long as the data (including the extension into the unobserved region) have low power on scales close to the Nyquist scale, we can expect the method described above to work.

By construction, the pure $E$ (resp. $B$) component will be an $E$ (resp. $B$) mode -- that is, the pure $E$ component will be a (sampled) derivative $\mathbf{D}_E$ of a band-limited potential, or equivalently it will be a discretization of a band-limited polarization field $\mathbf{p}_{pE}$ with $\mathbf{D}_B^\dag\cdot\mathbf{p}_{pE}=0$.  Similarly, the ambiguous component is by construction ambiguous, satisfying both $E$ and $B$ mode conditions.  If the method fails, therefore, it will do so via a lack of purity of the supposedly pure components.  This can be assessed by checking whether a map initially containing only $E$ modes has contamination in the pure $B$ component, and also by checking orthogonality of the three components

The above considerations apply to smooth data.  Of course, the assumption of smoothness does not apply to noise in the data, so the question of how the decomposition acts on the noise is a very important one. For both signal and noise, the only way to know if the method is working is to perform numerical tests.  Since the decomposition method is linear, we
can measure its treatment of signal and noise separately.

\begin{table}
\begin{center}
\begin{tabular}{|c|c|c|c|c|c|}
\hline
Parameters & $B_{\rm max}$ & $B_{\rm rms}$ & $\xi_{\rm sig}$& $\xi_{\rm noise}$ & $A_{\rm rms}^{\rm noise}$
\\
\hline
Fiducial & $2.8\times 10^{-3}$&$5.5\times 10^{-5}$&1.0045& 1.022 & 0.20\\
$P_B/P_E=0$&$2.5\times 10^{-4}$&$2.1\times 10^{-5}$&1.0045&1.21&0.39\\
$P_B/P_E=1$&$3.3\times 10^{-2}$&$4.2\times 10^{-4}$&1.0045&1.021&0.20\\
$P_B/P_E=100$&$6.5\times 10^{-2}$&$1.0\times 10^{-3}$&1.0045&1.021&0.19\\
$T_{\rm src}=5$&$8.0\times 10^{-3}$&$1.4\times 10^{-4}$&1.0045&1.019&0.19\\
$T_{\rm src}=2$&$2.6\times 10^{-2}$&$7.8\times 10^{-4}$&1.0060&1.018&0.19\\
$\Delta_{\rm src}=0$&$7.9\times 10^{-3}$&$1.5\times 10^{-4}$&1.0047&$1.018$&0.19\\
$N_{\rm sing}=100$&$0.10$&$9.3\times 10^{-2}$&1.020&1.28&0.51\\
$N_{\rm sing}=1800$&$7.5\times 10^{-4}$&$4.4\times 10^{-5}$&1.018&1.57&0.90\\
\hline
\end{tabular}
\caption{Results of tests of the E/B/A decomposition algorithm.  The quantities $B_{\rm rms}$ and $B_{\rm max}$ are the
maximum and rms contamination of the pure B component, which should ideally be zero, as a fraction of the input 
rms power.  The quantities $\xi_{\rm sig},\xi_{\rm noise}$, defined in equation (\ref{eq:orth}), quantify the orthogonality of the three components in the signal and noise map respectively. Finally, $A_{\rm rms}^{\rm noise}$ is the noise rms found in the ambiguous 
component, as a fraction of the input noise rms.  The first row shows results for the fiducial
parameters of Table \ref{table:params}.  In each subsequent row, one parameter is varied
from the fiducial values.}
\label{table:results}
\end{center}
\end{table}

The simulated map in Figure \ref{fig:extend} (hereinafter referred to as the signal map) will be used to illustrate the performance of the method.   The top panel of Figure \ref{fig:decomptest} shows the result of applying the E/B/A decomposition procedure to this map.  The decomposition method has a number of adjustable parameters, with the values listed in Table \ref{table:params}. 
The input map contained only $E$ modes, so the pure $B$ component should vanish.  Note that the amplitude of this component has been increased by a factor $10^4$ simply to make it visible. 

The bottom panel of Figure \ref{fig:decomptest}
shows an E/B/A decomposition of a map containing pure white noise, with the same observation
geometry.  In this case, one expects $E$ and $B$ to be comparable in amplitude, so the B component is not amplified.

To quantify the algorithm's performance, I assess the following: the extent of contamination of the pure $B$ mode by $E$ power, the orthogonality of the three components, and the fraction of noise power in the ambiguous mode.  
The top line of Table \ref{table:results} shows the results of these assessments.  
To quantify the contamination of the pure $B$ mode, I list both the maximum and
the rms polarization amplitudes
of the polarization in the pure $B$ map in Figure \ref{fig:decomptest}(a), compared to
the rms amplitude of the input map.  Since the contamination is highly concentrated near the boundary, the maximum is far larger than the rms, although it is still quite small.

To assess orthogonality of the pure $E$, pure $B$, and ambiguous components, I compute the total
polarization power in each of the three components ($P\equiv\sum_p (Q_p^2+U_p^2)$ over all pixels),
as well as for the input map.  I then compute
\begin{equation}
\xi = \sqrt{P_{pE}+P_{pB}+P_a\over P_{\rm input}},
\label{eq:orth}
\end{equation}
which equals 1 if the three components are orthogonal and exceeds 1 if they are positively
correlated.  This quantity is computed for either the signal map or the noise map.  The noise map provides a more stringent test, due to its abundance of high-frequency power.

\begin{figure*}
\epsfig{width=6in,figure=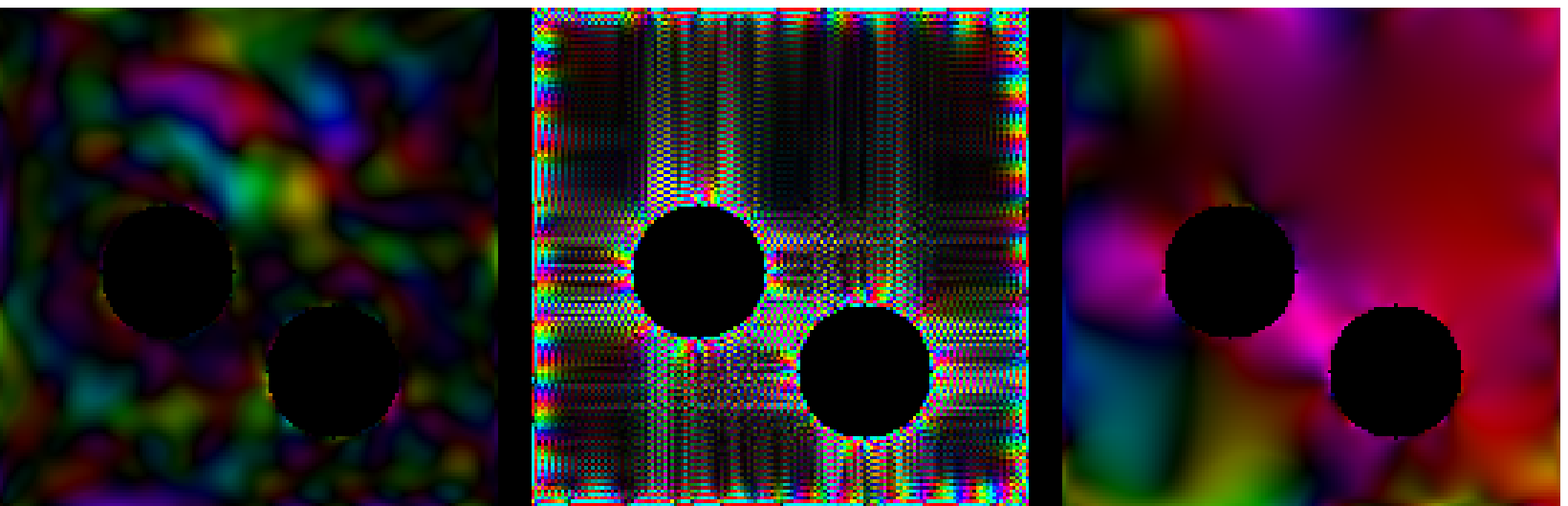}
\vskip 0.1in
\epsfig{width=6in,figure=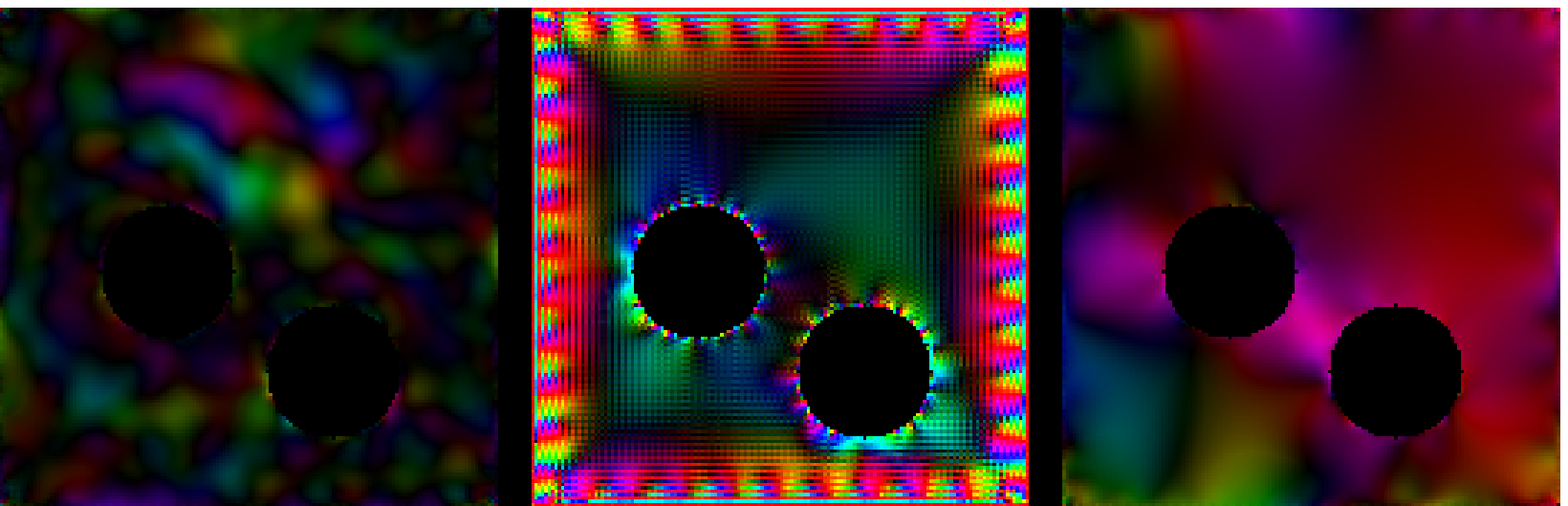}
\vskip0.1in
\epsfig{width=6in,figure=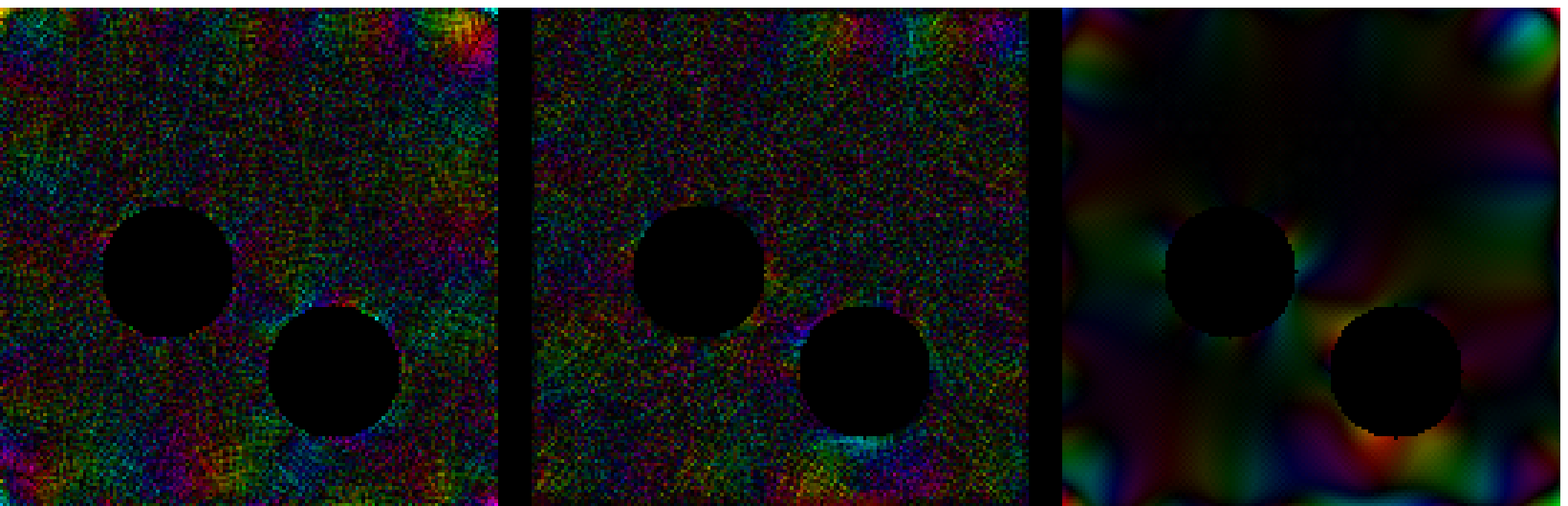}

\caption{Effects of varying decomposition parameters. (a) The power spectrum index
$n_s$ used for the extension is reduced from 4 to 2.  (b) The number of source points used
in finding biharmonic functions is reduced by setting the parameter $T_{\rm source}$ to 2.
(c) The number of singular values retained in finding biharmonic functions is reduced to 100.
Panels (a) and (b) show results for the simulated map of Figure \ref{fig:extend}, with the
pure B component enhanced by a factor $10^4$ as in Figure \ref{fig:decomptest}.  Panel (c)
is for a white noise map, with no B enhancement.}
\label{fig:varyparams}
\end{figure*}
Finally, the table lists the rms polarization amplitude in the ambiguous component of the noise map, relative
to the rms input noise power.  Once again, one could use either the simulated signal or the
noise map in this assessment.  The advantage of using the noise power is that we can predict
approximately what we expect to see from simple considerations.  The
number of ambiguous modes below the Nyquist frequency  is approximately equal to twice the length of the boundary in pixels \cite{BZT}.  The total number of modes that can be measured is equal to twice the
number of pixels.  In a white-noise map, all orthonormal modes should have equal power,
so the ratio of ambiguous-mode rms to total rms should be $\sqrt{N_{\rm bdy}/N_{\rm pix}}$.
The number of boundary pixels in our case is $N_{\rm bdy}=932$, and the number of observed
pixels is $N_{\rm pix}=19986$, so we expect the ratio to be approximately $0.2$.

The results show low levels of contaminated power and near-orthogonality as desired, and 
a level of ambiguous power in the noise map quite close to the theoretical estimate. 

The fiducial values in Table \ref{table:params} were chosen by trial and error to yield good results for these and similar tests, although they are certainly not the result of a systematic optimization
procedure.  On the contrary, increasing some parameters ($T_{\rm ext},T_{\rm source}$ in particular) causes
the test results continue to improve very modestly, at the cost of greater computation time.
The optimal choice of parameters will of course depend on the details of the data
set to be analyzed and on the tradeoff between computation time and accuracy.

Table \ref{table:results} and Figure \ref{fig:varyparams} show the results of varying some of these parameters.  For example, varying the spectral index used in extending the data to $n_s=2$
causes the extension to be less smooth, resulting in high-frequency power leaking into the 
pure B component [Figure \ref{fig:varyparams}(a)].  Reducing the number of source points
used to find the biharmonic functions also results in increased leakage [Figure \ref{fig:varyparams}(b).]  In both cases, recall that the B component is increased by $10^4$ in amplitude to make it visible: as Table \ref{table:results} indicates, the actual levels of contamination are still quite small.

The E/B/A decomposition algorithm treats E and B identically, except in the adoption of different
power spectrum normalizations for E and B modes used in extending the data.  The line in Table \ref{table:results}
showing the effect of adopting a power spectrum ratio $P_B/P_E=100$ illustrates this breaking
of symmetry: as expected, there is more leakage into the pure B mode when the extension is
heavily weighted towards B modes.  Equivalently, the fiducial ratio of 0.01 results in more leakage from $B$ into $E$ than vice versa.  The fiducial choice $P_B/P_E=0.01$ is based on the assumption that leakage from E into B is more of a concern than leakage from B into E.  If it
is deemed important to maintain E/B symmetry in the process, however, one can
 choose $P_B/P_E=1$.  The resulting rms leakage in this case is still quite less than one part
 in $10^3$, although the peak contamination, near the edges, rises to a few percent.

Some poor parameter choices result in numerical errors that can be seen in the failure
of the components to be orthogonal and in the excess power going into the ambiguous mode.  In particular, as noted in Section \ref{sec:extend}, forcing the B component power spectrum $P_B$ to be identically zero in the extension leads to numerical instability as the actual B modes are
shoehorned into E power.  Keeping too many singular values in solving for the biharmonic 
functions also leads to numerical instability, causing the ambiguous component to contain non-ambiguous modes [Figure \ref{fig:varyparams}(c)].

\section{Discussion}
\label{sec:discuss}

\begin{figure*}
\epsfig{width=6in,figure=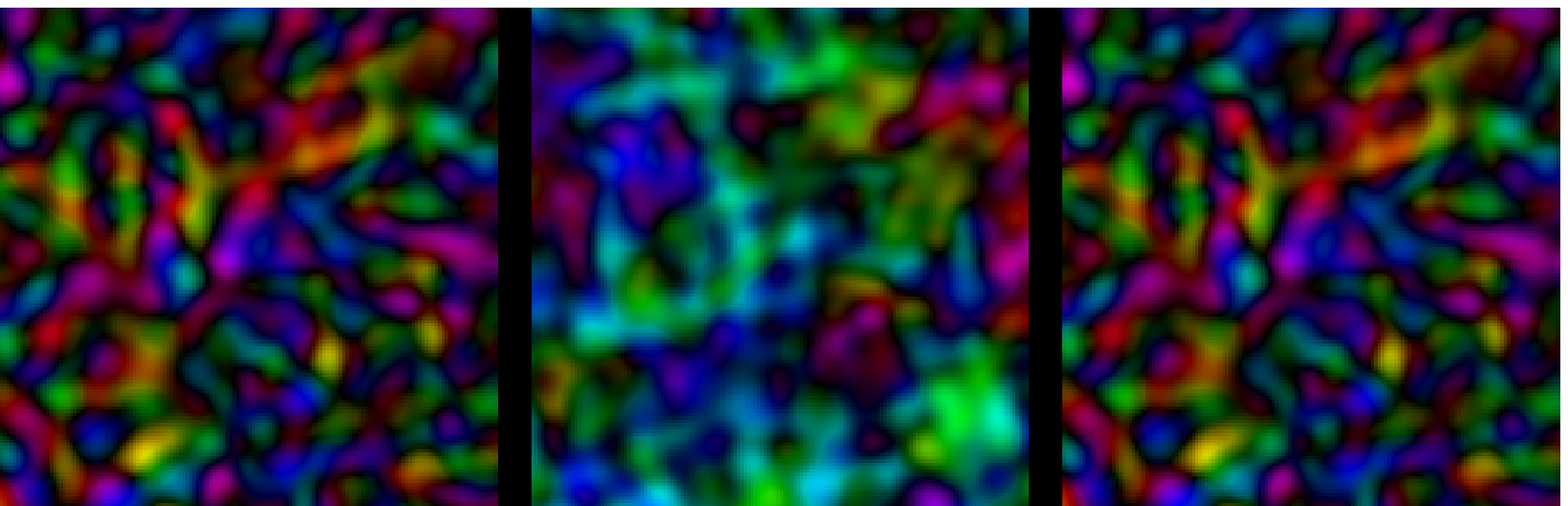}
\vskip0.1in
\epsfig{width=6in,figure=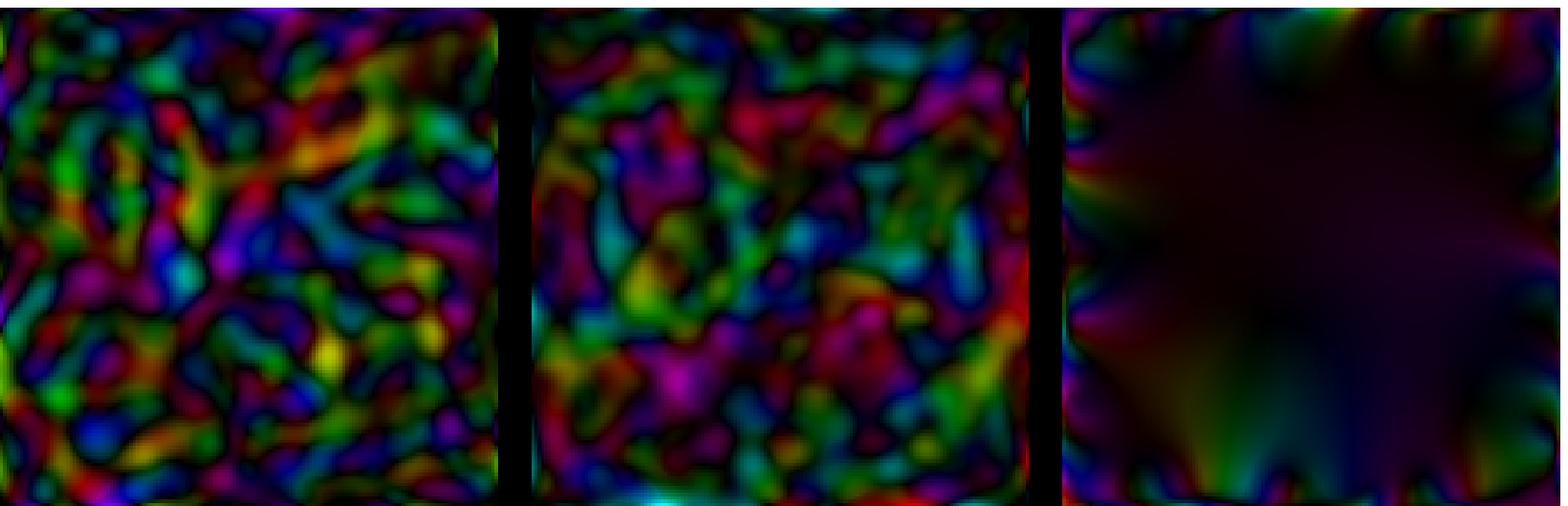}

\caption{Illustration of E/B/A decomposition. The upper panel
shows the $E$ component, $B$ component, and the sum of the two components
for a simulated map based on a $\Lambda$CDM cosmological model
with a tensor-scalar ratio $T/S=0.05$. The map contains $150\times 150$
pixels of size $3'$ and was smoothed with a $\sigma=12'$ beam. The
$B$ component is enhanced by a factor 15 for visibility.
The lower panel shows the result of the E/B/A decomposition: the pure $E$,
pure $B$, and ambiguous components are shown from
left to right. Again, the pure $B$ component is enhanced by
a factor 15.}
\label{fig:realistic}
\end{figure*}

The tests in the previous section indicate that the methods described in this paper can give a good approximation to the pure E, pure B,
and ambiguous modes of a CMB polarization data set.  The components are close to orthogonal,
and there is very low leakage of E modes into the pure B component.  The leakage that
does occur is, not surprisingly, close to the boundary of the observed region, where
effects of boundary discretization are most important.

Figure \ref{fig:realistic} provides an additional illustration of 
the method, using more realistic input data than the simple power-low
power spectra in previous examples. An input map was created containing
both $E$ and $B$ modes, with power spectra computed by CAMB \cite{camb}
based on the best-fit WMAP $\Lambda$CDM cosmology \cite{wmap7yrparams},
with a tensor-to-scalar ratio $T/S=0.05$. The simulated map is a $7.5^\circ
\times 7.5^\circ$ square, pixelized into $150\times 150$ pixels $3'$ in
size. The map was smoothed with a $\sigma=12'$ Gaussian beam. As
in the previous example, the map was simulated on a larger $1200\times 1200$
grid and truncated, so that it would not have periodic boundary conditions.
This map is sensitive to multipoles $30\lesssim l\lesssim 1000$.

The upper panel of Figure \ref{fig:realistic} shows the $E$ and $B$
components of the input map, as well as their sum. The 
lower panel shows the result of performing the E/B/A decomposition
on the sum.
In both cases, the $B$ component is enhanced by a factor 15 for
visibility.
The recovered $E$ and $B$ components look qualitatively similar to
the input components, especially away from the edges. The pure $E$
component contains 81\% of the input $E$ power ($Q^2+U^2$). The pure $B$
component contains 46\% of the input $B$ power. The ambiguous
component contains roughly 20\% of the
total power in the input map. As expected, nearly all of this ambiguous
power comes
from the $E$ component of the original map. The three components are
very close to orthogonal: the quantity $\xi$ in equation (\ref{eq:orth})
is $1.005$.

The method described in this paper was designed to avoid operations involving the solution of
$N_{\rm pix}$-dimensional linear systems.  The scaling of the various steps in the algorithm with
data size is therefore of interest.  Imagine a data set with $N_{\rm pix}$ pixels and a boundary
of length $N_{\rm bdy}$ pixels.  If the data set is reasonably round and has a smooth
boundary, then
we expect $N_{\rm bdy}\sim N_{\rm pix}^{1/2}$.  (If the data are extremely ``holey,'' due, e.g.,
to the removal of many point sources, then $N_{\rm bdy}$ might be much larger.  One might
wish to search for special
techniques for treating this particular case of many small round holes in the data.)

The smooth extension of the data involves the solution
of a linear system of size  $N_{\rm cons}\sim (T_{\rm ext}N_{\rm bdy})$, so the scaling of this
step is $O((T_{\rm ext}N_{\rm bdy})^3)\sim O( T_{\rm ext}^3N_{\rm pix}^{3/2})$, assuming a ``nice'' 
boundary.  The most expensive step in the solution of this system is a Cholesky decomposition,
which depends only on the pixel geometry and not on the data itself.  Thus if multiple maps with
the same geometry are to be analyzed (e.g., in Monte Carlo simulations), this step can be
precomputed.  The parts of the smooth extension that cannot be precomputed scale at most as
$O(N_{\rm cons}^2)$.  (In any case, the method I have described for smooth extension is hardly unique;
it is easy to imagine that faster ones can be found.)

The step involving the solution of the bilaplacian equation scales similarly: there is a
precomputable singular value decomposition scaling as $O(N_{\rm bdy}^2N_{\rm src})\sim
O(T_{\rm src}N_{\rm bdy}^3)\sim O(T_{\rm src}N_{\rm pix}^{3/2})$, again assuming a ``nice'' boundary for the last step.  The non-precomputable part of the process scales as 
$O(T_{\rm src}N_{\rm bdy}^2)\sim O(T_{\rm src}N_{\rm pix})$.

The rest of the process involves Fourier transforms, which of course scale as $O(N_{\rm pix}\ln
N_{\rm pix})$.  

I have described and implemented the algorithm in the flat-sky approximation for simplicity.
Each step in the process generalizes in a perfectly natural way to the spherical sky, so a spherical implementation should be perfectly possible.  In this case, the Fourier transforms must be
replaced with spherical harmonic transforms.   The portions of the
algorithm that scale as $N_{\rm pix}\ln N_{\rm pix}$ will then scale as $N_{\rm pix}^{3/2}$ (i.e.,
the same as the HEALPix \cite{healpix} programs {\tt synfast} and {\tt anafast}, and still at least as fast as the slowest other steps in the algorithm).

Alternative methods of performing an E/B/A decomposition have been proposed \cite{kimnaselsky,zhao,caofang}.  These methods are all potentially useful, but they differ from the one presented here in important ways. 
Some methods \cite{kimnaselsky,zhao} involve finding scalar-valued derivatives of the pure components (essentially, $\mathbf{D}_{E,B}^\dag\cdot\mathbf{p}$, but do not yield the actual polarization maps (i.e., Stokes $Q,U$) of the components.  This has the advantage that the ambiguous contribution to the scalar maps is more concentrated at the boundary, so that approximate purification can be achieved by removing data near the edges.  However, these methods do not allow for purification of the actual observables ($Q,U$), as for these quantities  the ambiguous modes persist far into the interior (Figure \ref{fig:decomptest}).  For some purposes, one may wish to analyze the pure $E$ and $B$ components of the actual polarization map.

In addition, there is a potentially promising method based on a wavelet decomposition \cite{caofang}.   It too involves removing the ambiguous modes via a hard cutoff near the boundary, but the cutoff is set in terms of the scale of each wavelet, rather than being a fixed number of pixels.  This is a sensible approach, as the distance an ambiguous mode persists into the interior of a map depends on the frequency of the source function on the boundary.  In contrast, the method I have described does not make any a priori assumption about the ambiguous modes being restricted to the proximity of the border.  Rather, it solves the relevant equation to determine how far from the border the ambiguous modes persist.

In preparing for the analysis of any particular data set, it would be extremely interesting to perform simulations to compare the performance of the various methods in detail.

\section*{Acknowledgments}

This work was supported by NSF awards 0507395 and 0922748.  I thank the Laboratoire
Astroparticule et Cosmologie at the Universit\'e Paris VII for hospitality while some of this
work was performed.  

\bibliography{purify}

\end{document}